\documentstyle[twocolumn,aps,prl,epsfig]{revtex}
\begin{document}
\draft
\title{Spin dynamics of strongly-doped La$_{1-x}$Sr$_x$MnO$_3$}
\author{L. Vasiliu-Doloc, J. W. Lynn,} 
\address{NIST Center for Neutron Research, National Institute of 
Standards and Technology, Gaithersburg, Maryland 20899 and \\
Center for Superconductivity Research, University of Maryland, 
College Park, MD 20742}
\author{Y.M. Mukovskii, A.A. Arsenov, D.A. Shulyatev}
\address{Moscow Steel and Alloy Institute, Moscow 117936, Russia}
%\date{\today  }
\maketitle

\begin{abstract} 

Cold neutron triple-axis measurements have been used to investigate the
nature of the long-wavelength spin dynamics in strongly-doped 
La$_{1-x}$Sr$_{x}$MnO$_3$ single crystals with $x$=0.2 and 0.3. 
Both systems behave like isotropic ferromagnets
at low $T$, with a gapless ($E_0 < 0.02$ meV) quadratic dispersion relation 
$E = E_0 + Dq^2$. The values of the spin-wave stiffness constant $D$ are large 
($D_{T=0}$ = 166.77 meV$\cdot \AA^2$ for $x$=0.2 and D$_{T=0}$ = 175.87 
meV$\cdot \AA^2$ 
for $x$=0.3), which directly shows that the electron transfer energy for the $d$
band is large. $D$ exhibits a power law behavior as a function of
temperature, and appears to collapse as $T \to T_C$. Nevertheless,
an anomalously strong quasielastic central component develops and
dominates the fluctuation spectrum as $T \to T_C$. Bragg scattering indicates 
that the magnetization near $T_C$ exhibits power law behavior, with $\beta  
\simeq 0.30$ for both systems, as expected for a 
three-dimensional ferromagnet.
 
\end{abstract}

{\bf I. INTRODUCTION}\\

Since the recent discovery of unusually large magnetoresistive effects in 
perovskite manganites, the doped LaMnO$_3$ class of materials$^1$ has generated 
continued interest and has motivated experimental and 
theoretical work devoted to understanding of the origin of this colossal 
magnetoresistance (CMR) phenomenon. The large variation in the 
carrier mobility originates from an insulator-metal transition that is closely 
associated with the magnetic ordering. The on-site exchange interaction between
the spins on the manganese ions is believed to be strong enough to completely 
polarize the ($e_g$) conduction electrons in the ground state, forming a 
``half-metallic" ferromagnet. However, hopping, and hence conduction, may only 
occur if the Mn core spins (formed  by the $d$ electrons in a $t_{2g}$ orbital)
on adjacent sites are parallel, which then directly couples ferromagnetic order 
with the 
electrical conductivity at elevated temperatures. This mechanism, known as the
double exchange mechanism,$^2$ was first proposed in the 1950s, and has provided
a good description of the evolution of the magnetic properties with band 
filling. However,
in order to fully explain all the properties of the CMR materials, strong
electron correlations,$^3$ and/or a strong electron-lattice coupling$^4$ in 
different polaronic approaches are invoked. 
Cooperative Jahn-Teller (JT) distortions associated with the Mn$^{3+}$ JT ions
have been evidenced from structural studies at low doping, where the system is 
insulating and antiferromagnetic, and may be an important contribution to 
orbital ordering, double exchange, and related spin ordering and transport 
properties observed at higher concentrations. As the doping concentration $x$ 
increases, the static JT distortion weakens progressively and the system becomes
metallic and ferromagnetic, with the CMR property observed for doping levels
$x > 0.17$. It is believed that in the absence of a cooperative effect in this
regime, local JT distortions persist on short time and length scales. These
short-range correlations would contribute, together with the electron 
correlations, to create an effective carrier mass necessary for large 
magnetoresistance. This unique class of half-metallic 
ferromagnets provides an excellent opportunity to elucidate the influence 
of such correlations on the lattice and spin dynamics, which can best be
probed by inelastic neutron scattering.

In the optimally doped regime with $x \sim 0.3$ it has been shown that the 
ground state spin
dynamics is essentially that expected for a conventional metallic ferromagnet
described by an isotropic Heisenberg model$^{5-7}$. For the 
Ca-doped system, however, results obtained on polycrystalline samples$^{8}$ 
have indicated a possible coexistence of spin-wave excitations and spin 
diffusion in the ferromagnetic phase. In particular, it was suggested that the 
quasielastic component of the scattering that develops rapidly as the Curie 
temperature is approached is associated with the localization of the $e_g$ 
electrons on the Mn$^{3+}$/Mn$^{4+}$ lattice, and may be related to the 
formation of spin polarons in the system$^9$. Furthermore, it is this spin 
diffusion that drives the ferromagnetic phase transition rather than the thermal
population of conventional spin waves.
In the present publication we report diffraction and inelastic measurements of 
the spin dynamics in the metallic ferromagnets La$_{0.8}$Sr$_{0.2}$MnO$_3$ and
La$_{0.7}$Sr$_{0.3}$MnO$_3$.\\

{\bf II. EXPERIMENT}\\

The single crystals used in the present neutron scattering experiments were 
grown at the Steel and Alloys Institute in Moscow, using the floating zone 
method. The crystals weighed 2.25 and 4.25 g, respectively. The samples 
were oriented such that the [100] and [010] axes of the rhombohedral $R\bar{3}c$
cell lie in the scattering plane. The neutron scattering measurements have been 
carried out on the NG-5 (SPINS) cold
neutron triple-axis spectrometer at the NIST research reactor.
The (002) reflection of pyrolytic graphite (PG) was used as monochromator and 
analyser for measuring the low-energy part of the spin-wave spectrum. 
We have used a flat analyzer with a fixed final energy E$_f$ = 3.7 meV, a cold 
Be filter on the incident beam,
and collimations 40$^{\prime}$-S-40$^{\prime}$-130$^{\prime}$ in sequence from 
the neutron guide to detector. This configuration offered an energy resolution 
of $\sim$ 0.15 meV, together with good q-resolution.
Each sample was placed in a helium-filled aluminum cell in a displex 
refrigerator. The sample temperature ranged from 15 to 325 K for 
La$_{0.8}$Sr$_{0.2}$MnO$_3$, and from 30 to 375 K for 
La$_{0.7}$Sr$_{0.3}$MnO$_3$, and was controlled to within 0.1$^o$.
                                                                       
The crystal structure of both systems at room temperature and below is 
rhombohedral ($R\bar{3}c$), with $a_0 \simeq b_0 \simeq c_0 \simeq 3.892$ $\AA$ 
for $x=0.2$ and $a_0 \simeq b_0 \simeq c_0 \simeq 3.884$ $\AA$ for $x=0.3$. \\

{\bf III. RESULTS AND DISCUSSION}\\

Figure 1 shows the integrated intensity of the (100) Bragg reflection as a 
function of temperature for both samples. This reflection has a finite nuclear 
structure factor, and therefore the intensity in the paramagnetic phase is 
nonzero. The increase in intensity below $T_C$ is due to magnetic scattering 
produced by the ferromagnetism of spins aligning on the manganese ions and
yielding a magnetic structure factor. The solid curve is a fit of the points
near $T_C$ to a power law. The best fits give $T_C$ = 305.1 K and a critical 
exponent $\beta$ = 0.29 $\pm$ 0.01 for La$_{0.8}$Sr$_{0.2}$MnO$_3$, and $T_C$ 
= 350.8 K and $\beta$ = 0.30 $\pm$ 0.02 for La$_{0.7}$Sr$_{0.3}$MnO$_3$. Both 
values of the critical exponent are slightly below, but rather close to, the 
well known three-dimensional Heisenberg ferromagnet model value of $\sim 1/3$.

We have investigated the spin dynamics in the (1,0,0) high-symmetry direction
in both samples. The ground state spin dynamics for a half-metallic ferromagnet
was not expected to differ much from the conventional picture of well defined 
spin waves, and we found that the long wavelength magnetic excitations were in
fact the usual spin waves, with a dispersion relation given by $E = E_0 + Dq^2$,
where $E_0$ represents the spin wave energy gap and the spin stiffness 
coefficient is directly related to the exchange interactions. The spin-wave
gap $E_0$ was too small to be measured directly in energy scans at the zone 
center, but very high-resolution measurements on the NG-5 (SPINS) cold-neutron
triple-axis spectrometer have allowed us to determine that $E_0 < 0.02$ meV for
both systems, which demonstrates that these are "soft" isotropic ferromagnets.
A previously reported value of $E_0$ = 0.75 meV for the $x$=0.3 system$^6$ was 
obtained 
from an extrapolation of higher $q$ data, not from direct high-resolution 
measurements as in the present case. The low-temperature values of
the spin-wave stiffness constant $D$ are large: $D_{T=0} = (166.77 \pm 1.51$) 
meV$\cdot \AA^2$ for $x$=0.2 and $D_{T=0} = (175.87 \pm 5.00$) meV$\cdot \AA^2$ 
for $x$=0.3, and show that the electron transfer energy for the $d$ band is 
large. The low temperature value of the spin stiffness constant gives a ratio
$D/k_BT_C \sim 6.34$ and 5.82 for the $x$=0.2 and 0.3 systems, respectively.
Both values are quite large, as might be expected for an itinerant electron
system. 

Figure 2 plots the temperature dependence of the spin-wave stiffness 
$D$. The data have
been analysed in terms of two-spin-wave interactions in a Heisenberg ferromagnet
within the Dyson formalism,$^{10}$ which predicts that the dynamical
interaction between the spin waves gives, to leading order, a temperature
dependence: 
\begin{equation}
D(T) = D_{0} \left\{1 - \frac{v_0 \overline{l^2} \pi}{S} \left(
\frac{k_B T}{4\pi D_{0}} \right)^{5/2} \zeta(\frac{5}{2}) \right\} ,
\label{T5/2}
\end{equation}
where $v_0$ is the volume of the unit cell, $S$ is the average value of the
manganese spin, and $\zeta (\frac{5}{2})$ is the Riemann zeta integral.
$\overline{l^2}$ is the moment defined by
$\overline{l^n} = \frac{S}{3D} \left\{ \sum l^{n+2} J({\bf l}) \right\}$
and which, compared to the square of the lattice parameter $a^2$, gives
information about the range of the exchange interaction. The solid curves
in Fig. 2 are fits to Eq.~\ref{T5/2}, and are in good agreement
with the experimental data for reduced temperatures $t = (T - T_C)/T_C$ up
to $t_1 \simeq$ -0.1 for La$_{0.8}$Sr$_{0.2}$MnO$_3$ and -0.14 for 
La$_{0.7}$Sr$_{0.3}$MnO$_3$. The 
fitted values of $\overline{l^2}$ give $\sqrt{\overline{l^2}}$ = (3.92 
$\pm 1.04)a_0$ for $x$=0.2, and $\sqrt{\overline{l^2}}$ = (3.84 $\pm 1.22)a_0$ 
for $x$=0.3, which indicates that the exchange interaction extends beyond
nearest neighbors in both systems. For $t > t_1$ the experimentally measured 
values of $D$ depart from the T$^{5/2}$ dependence, having rather a power law 
behavior and 
appearing to collapse as T $\to$ T$_c$. The dashed lines in Fig. 2 are fits to
a power law $\left [1 - {T \over T_C}\right ]^{\nu^\prime - \beta}$, where
$\nu^{\prime}$ is the critical exponent for a three-dimensional ferromagnet. 

In the course of our measurements we have noticed that the central peak has a
strong temperature dependence on approaching $T_C$, while typically the central 
peak 
originates from  weak temperature-independent nuclear incoherent scattering.
Figure 3(a) shows two magnetic inelastic spectra collected at 300 and 325 K, and
reduced wave vector $q$ = 0.035 away from the (100) reciprocal point in the
La$_{0.7}$Sr$_{0.3}$MnO$_3$ ($T_C$ = 351 K). A flat background of 4.9 counts 
plus an elastic incoherent nuclear peak of 110 counts, measured at 30 K, have 
been subtracted from these data. We can clearly see the
development of the quasielastic component, comparable in intensity to the spin
waves, and the temperature dependence of the strength of this scattering 
is shown in Fig. 3(b) as a function of temperature.  We observe a significant
intensity starting at 250 K ($\sim$ 100 K below $T_C$), and the scattering peaks
at $T_C$. At and above $T_C$ all the scattering is quasielastic. 
For typical isotropic
ferromagnets, such as Ni, Co, Fe, any quasielastic scattering below T$_C$ is
too weak and broad to be observed directly in the data, and can only be
distinguished by the use of polarized neutron techniques. In Fig. 3(a)
we can nevertheless see that the spectrum starts to be dominated by this
quasielastic
component at temperatures well below T$_C$. The appearance in the ferromagnetic
phase of a quasielastic component 
was first observed on Ca-doped polycrystalline samples,$^8$ and it has 
been suggested that it is associated with the localization of the $e_g$ 
electrons on the Mn$^{3+}$/Mn$^{4+}$ lattice, and may be related to the 
formation of spin polarons in the system.$^9$ 
We have observed a similar anomalous behavior of the central peak in the more
lightly-doped system La$_{0.85}$Sr$_{0.15}$MnO$_3$,$^{11}$ but for that doping
we find that the central component 
becomes evident only much closer ($\sim$ 25 K) to the Curie temperature.
Similar data have been obtained on both polycrystalline and single crystal
samples of the Ba-doped system.$^{12}$
It thus appears that the coexistence of spin-wave excitations and spin
diffusion is a common characteristic for many perovskite manganites, and that it
may be relevant for the giant magnetoresistance property of these systems.
It is therefore important to pursue the study of this aspect with polarized
neutron techniques, in order to determine the nature of the fluctuations
involved in this new quasielastic component to the fluctuation spectrum.

Research at the University of Maryland is supported by the NSF under Grant
DMR 97-01339 and by the NSF-MRSEC, DMR 96-32521. Experiments on the NG-5
spectrometer at the NIST Research Reactor are supported by the NSF under
Agreement No. DMR 94-23101.

\newpage

\parindent 0cm

$^1$G.H. Jonker and J.H. van Santen, Physica {\bf 16}, 337 
(1950); E.O. Wollan and W.C. Koehler, Phys. Rev. {\bf 100}, 545
(1955); G.H. Jonker, Physica {\bf 22}, 707 (1956).

$^2$C. Zener, Phys. Rev. {\bf 82}, 403 (1951); P.W. Anderson and H. 
Hasegawa, Phys. Rev. {\bf 100}, 675 (1955); P.G. de Gennes, Phys. Rev. 
{\bf 100}, 564 (1955).

$^3$Y. Tokura, A. Urushibara, Y. Moritomo, T. Arima, A. Asamitsu, G. Kido, and 
N. Furukawa, J. Phys. Soc. Jpn. {\bf 63}, 3931 (1994).

$^{4}$A.J. Millis, P.B. Littlewood, and B.I. Shraiman, Phys. Rev.
Lett. {\bf 74}, 5144 (1995); A.J. Millis, Phys. Rev. B {\bf 55}, 6405 (1997).

$^{5}$T.G. Perring, G. Aeppli, S.M. Hayden, S.A. Carter, J.P. Remeika, and
S.-W. Cheong, Phys. Rev. Lett. {\bf 77}, 711 (1996).

$^{6}$M.C. Martin, G. Shirane, Y. Endoh, K. Hirota, Y. Moritomo, and Y. Tokura,
Phys. Rev. B {\bf 53}, 14285 (1996).

$^{7}$A.H. Moudden, L. Pinsard, L. Vasiliu-Doloc, A. Revcolevschi, Czech. J. 
Phys. {\bf 46}, 2163 (1996).

$^{8}$J. W. Lynn, R.W. Erwin, J.A. Borchers, Q. Huang, and A. Santoro,
Phys. Rev. Lett. {\bf 76}, 4046 (1996).

$^9$J.W. Lynn, R.W. Erwin, J.A. Borchers, A. Santoro, Q. Huang, J.-L. Peng, R.L.
Greene, J. Appl. Phys. {\bf 81}, 5488 (1997).

$^{10}$D.C. Mattis, {\em The theory of magnetism}, Spinger-Verlag, Heidelberg, 
1981.

$^{11}$L. Vasiliu-Doloc, J.W. Lynn, A.H. Moudden, A.M. de Leon-Guevara, A. 
Revcolevschi, J. Appl. Phys. {\bf 81}, 5491 (1997).

$^{12}$J.W. Lynn, L. Vasiliu-Doloc, S. Skanthakumar, S.N. Barilo, G.L. Bychkov
and L.A. Kurnevitch, private communication.

\newpage

\begin{center}

{\bf FIGURE CAPTIONS}

\end{center}

FIG. 1. Temperature dependence of the integrated intensity of the (100) Bragg 
peak for (a) La$_{0.8}$Sr$_{0.2}$MnO$_3$ and (b) La$_{0.7}$Sr$_{0.3}$MnO$_3$. 
There is a
nuclear contribution to this peak, and the additional temperature-dependent 
intensity originates from the onset of the ferromagnetic order at T$_C$ = 305 K
for the $x$=0.2 system, and $T_C$ = 350.8 K for $x$=0.3.
The solid curves are fits of the points near $T_C$ to a power law. \\

FIG. 2. Spin-wave stiffness coefficient $D$ in $E = E_0 + Dq^2$ as a 
function of temperature for (a) La$_{0.8}$Sr$_{0.2}$MnO$_3$ and (b) 
La$_{0.7}$Sr$_{0.3}$MnO$_3$. The solid curves are fits to Eq. (1). $D$ appears 
to vanish at the ferromagnetic transition temperature, as expected for a 
conventional ferromagnet. The dashed curves are fits to a power law. \\

FIG. 3. (a) Constant-{\bf q} magnetic inelastic spectra collected at 300 and 
325 K and a reduced wave vector vector q = (0, 0, 0.035) for 
La$_{0.7}$Sr$_{0.3}$MnO$_3$ ($T_C$ = 350.8 K), and (b) temperature dependence 
of the integrated intensity of the quasielastic central component. 
The dominant effect
is the development of a strong quasielastic component in the spectrum. Above 
$T_C$, all the scattering in this range of q is quasielastic. 

\newpage
\clearpage
\widetext

\vspace*{3cm}
\begin{center}
\epsfig{file=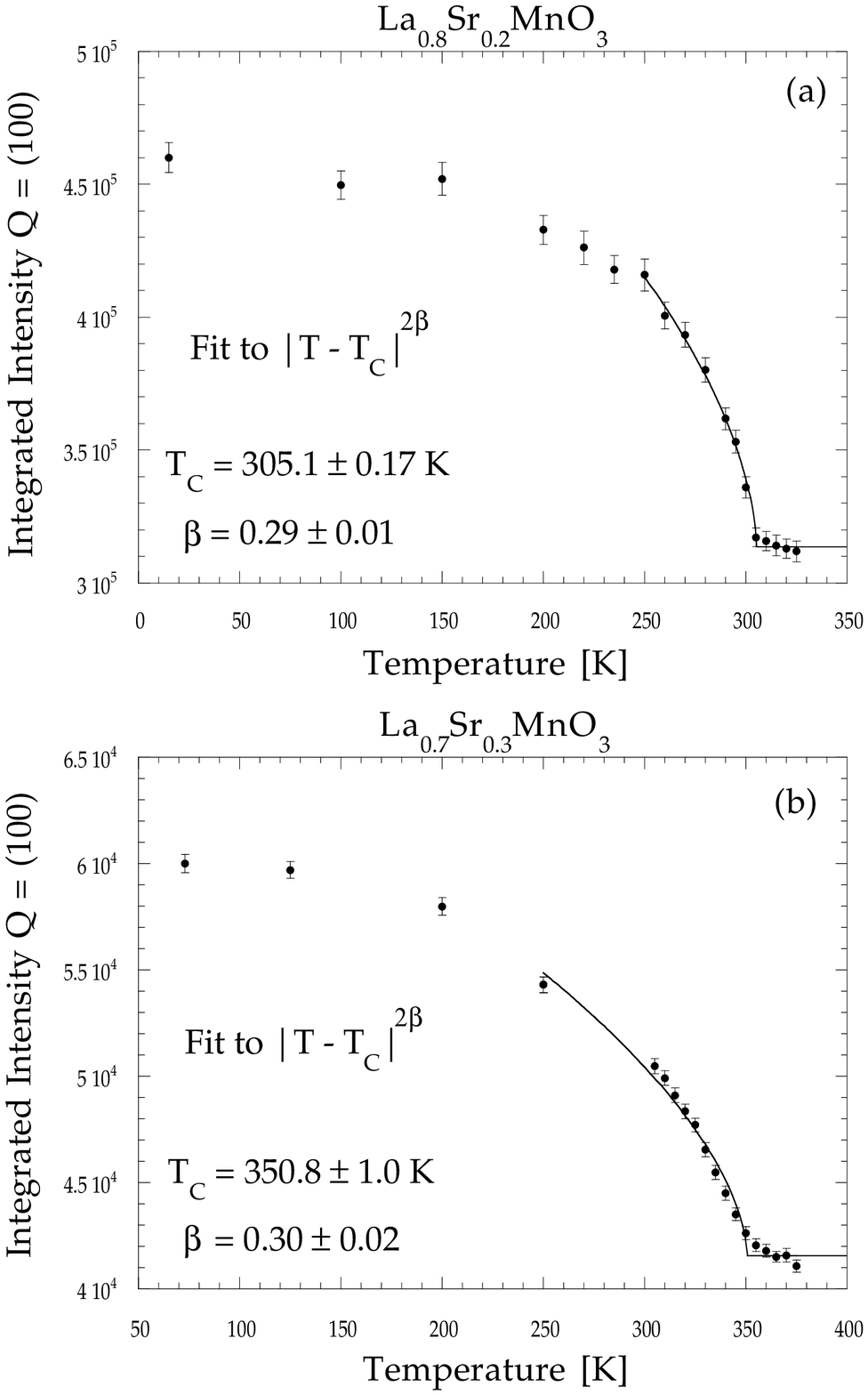,width=12.5truecm}
\end{center}

\vspace{1.5cm}
\large
\makebox[8cm]{   }\makebox[1.8cm][l]{Fig. 1:}\makebox[7cm][l]{L. Vasiliu-Doloc 
et al.}
%\makebox[8cm]{   }\makebox[1.8cm][l]{  }\makebox[7cm][l]{J. Appl. Phys.}

\newpage
\clearpage
\vspace*{3cm}
\begin{center}
\epsfig{file=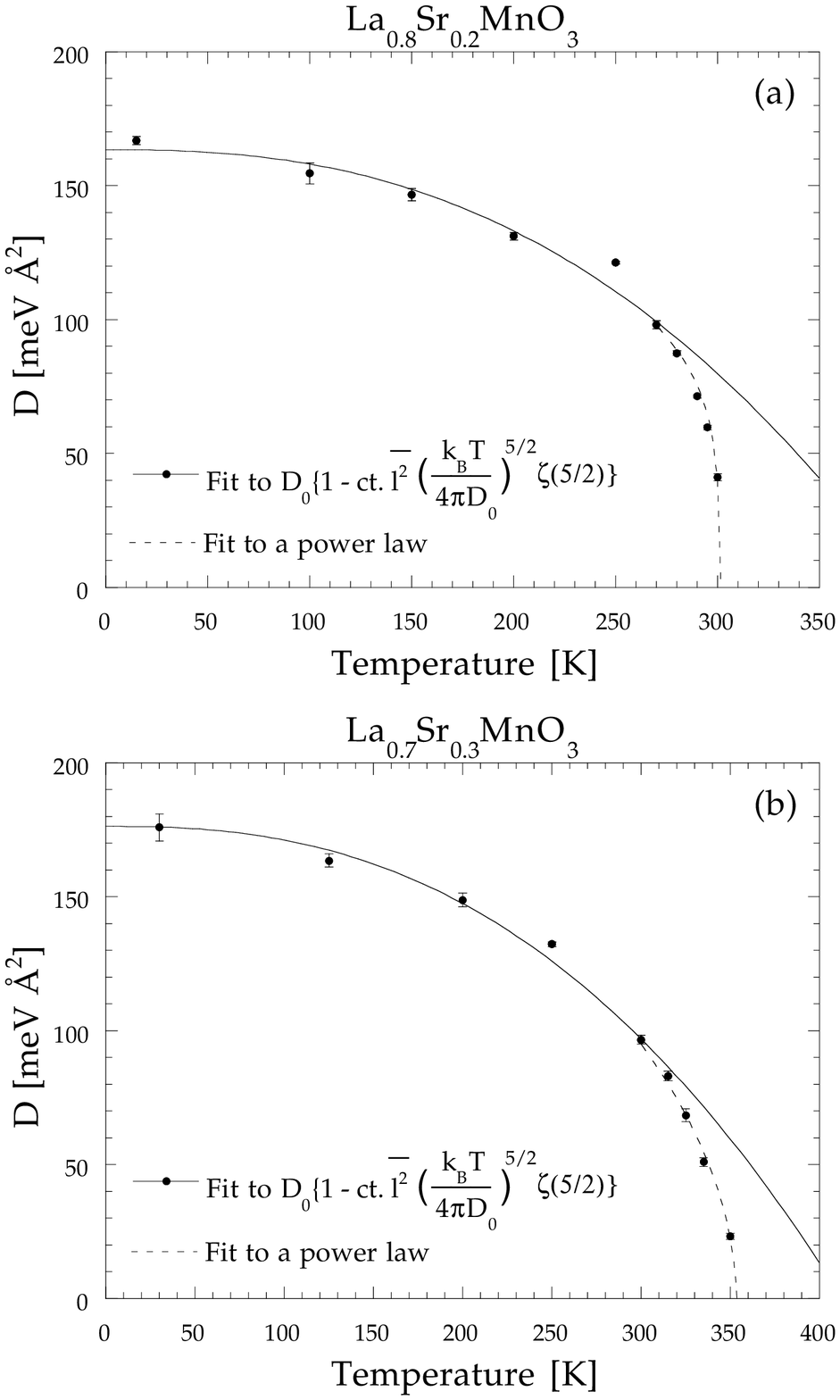,width=12.5truecm}
\end{center}

\vspace{1.5cm}
\large
\makebox[8cm]{   }\makebox[1.8cm][l]{Fig. 2:}\makebox[7cm][l]{L. Vasiliu-Doloc 
et al.}
%\makebox[8cm]{   }\makebox[1.8cm][l]{  }\makebox[7cm][l]{J. Appl. Phys.}

\newpage
\clearpage
\vspace*{3cm}
\begin{center}
\epsfig{file=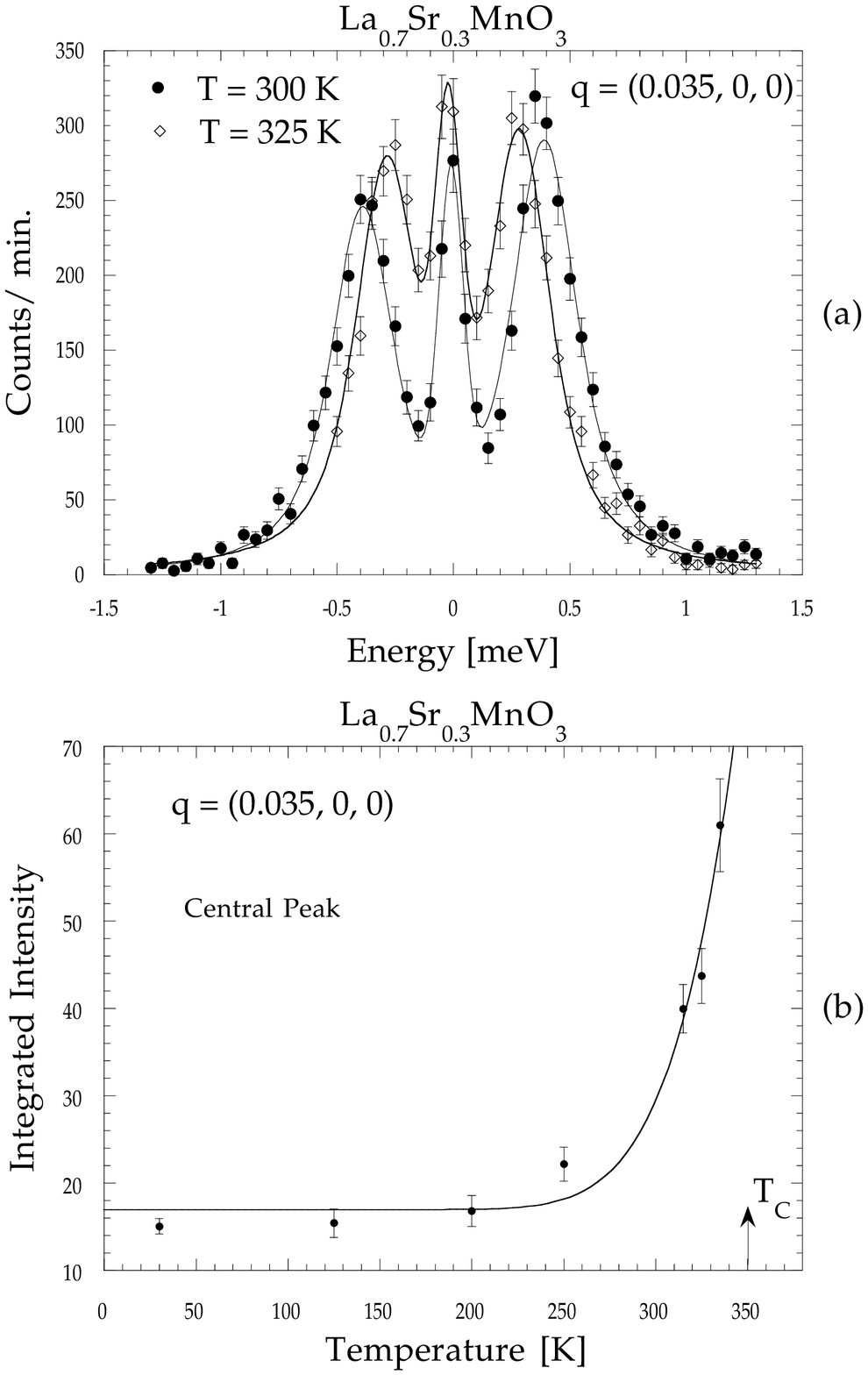,width=12.5truecm}
\end{center}

\vspace{1.5cm}
\large
\makebox[8cm]{   }\makebox[1.8cm][l]{Fig. 3:}\makebox[7cm][l]{L. Vasiliu-Doloc 
et al.}
%\makebox[8cm]{   }\makebox[1.8cm][l]{  }\makebox[7cm][l]{J. Appl. Phys.}

\end{document}